\documentclass[aps,pra,showpacs,notitlepage]{revtex4-1}

\usepackage{graphicx}

\begin{document}

\title{Direct evaluation of measurement uncertainties\\ by feedback compensation of decoherence}

\author{Holger F. Hofmann}
\email{hofmann@hiroshima-u.ac.jp}
\affiliation{
Graduate School of Advanced Science and Engineering, Hiroshima University,
Kagamiyama 1-3-1, Higashi Hiroshima 739-8530, Japan
}

\begin{abstract}
It is shown that measurement uncertainties can be observed directly by evaluating the feedback compensation of the decoherence induced by the measured system on a probe qubit in a weak interaction occuring between state preparation and measurement. The uncompensated decoherence is described by the measurement uncertainties introduced by Ozawa in Phys. Rev. A 67, 042105 (2003), confirming the empirical validity of measurement theories that combine the initial information of the input state with the additional information provided by each measurement outcome. 
\end{abstract}

\maketitle
As new quantum technologies are being developed, fundamental questions may obtain new and unexpected practical significance. An interesting question concerns the measurement error associated with the observation of a physical property in an uncertainty limited measurement \cite{Hei27,Bus07,Rod19}. Since it is not possible to go back in time to perform a precise measurement of the target observable, the uncertainty principle itself seems to prevent any observable effects of the measurement error. A mathematical definition of the error based on the representation of values by their corresponding operators was proposed by Ozawa \cite{Oza03}, but this mathematical definition has been criticized precisely because it refers to hypothetical properties that do not appear in the observable statistics of quantum states \cite{Wat11,Bus13,Dre14,Bus14,Roz15}. In fact, the insistence on concepts considered to be ``useful'' in quantum information protocols might have done more harm than good in the objective and scientific discussion of the issue. Specifically, the consistency of Ozawa's theory with the results of weak measurements and the fact that the results of error free measurements can be anomalous weak values have not been sufficiently recognized as convincing evidence in favor of either Ozawa's theory or the theory of weak values, possibly because it is suspected that both theories might be misrepresentations of quantum interference effects  \cite{Lun10,Lee16,Iin16,The18,Dre15,Sok16}. It is therefore extremely important to consider the possibility that some practical effects associated with a physical property between state preparation and measurement might have been overlooked. In the following, it will be shown that this is indeed the case: Ozawa's uncertainties describe the experimentally observable fluctuations of weak forces in a quantum interaction, with the optimal estimate given by the corresponding weak values.  

Taking inspiration from recent implementations of quantum feedback protocols \cite{Vij12,Bol14,Soa14,Wak17,Mav17,Nag20}, the essential idea is to use the outcome of a measurement as a feedback signal on a quantum probe that weakly interacted with the system between state preparation and measurement. In this case, the uncertainty of the target observable causes a small but detectable amount of decoherence in the probe system. This decoherence corresponds to a random unitary operation, the parameter of which is determined by the value of the uncertain observable. If more information on that observable is obtained in a quantum measurement, this information can be used to implement a negative feedback to compensate the decoherence suffered by the probe. It is then possible to directly observe the uncertainty of the quantum measurement in the amount of uncompensated decoherence of the probe. A standardized setup for feedback compensation of decoherence using a qubit as a probe can thus be used to implement an operational definition of measurement uncertainty. The measurement uncertainty determined in this standardized manner is a technically relevant property of the quantum measurement that provides practical information on the performance of the measurement within a larger quantum circuit. The method proposed here thus provides an objective benchmark test of measurement uncertainties, independent of any additional theoretical assumptions.

In the following analysis, I show that the uncompensated decoherence in the feedback compensation scenario described above is given by the Ozawa uncertainty of the measurement. The optimal definition of the measurement outcomes is given by the weak values of the target observable for each measurement outcome \cite{Lun10,Iin16,The18}. Complete feedback compensation is possible when the weak values of pure state inputs are all positive and real. Weak values thus provide an accurate representation of quantum fluctuations in the absence of direct measurements of the fluctuating observable. The empirical approach introduced in the present paper demonstrates not only that Ozawa uncertainties correctly describe the fluctuations of weak forces associated with the target observable, it also shows that weak values provide the optimal estimate of these weak forces, revealing that the dependence of weak values on the measurement context is a fundamental characteristic of all quantum fluctuations \cite{Tol07,Hof20}.

\begin{figure}[th]
\begin{picture}(500,360)
\put(50,0){\makebox(360,360){
\scalebox{0.8}[0.8]{
\includegraphics{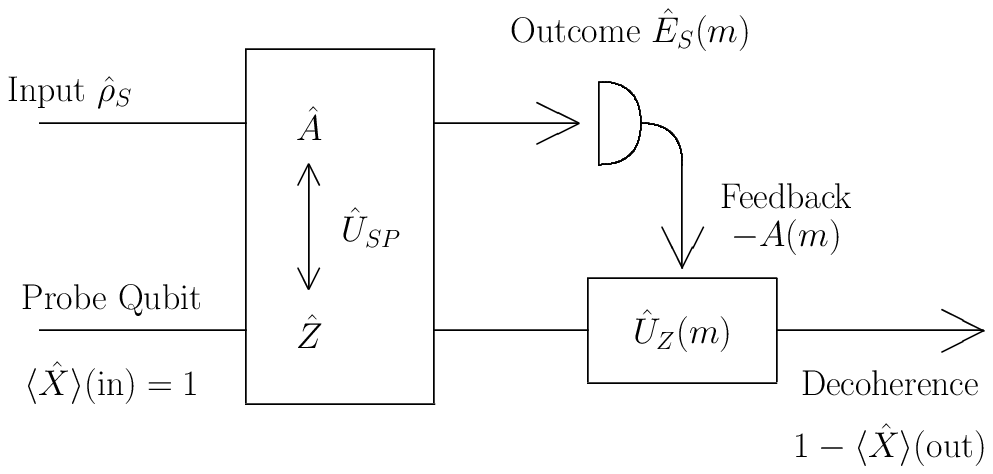}}}}
\end{picture}
\vspace{-5cm}
\caption{\label{fig1}
Feedback compensation of the decoherence in a probe qubit caused by a weak interaction with the noisy property $\hat{A}$ of a quantum system. The value of the property $\hat{A}$ is estimated based on the measurement outcome $m$ and a corresponding negative feedback of $-A(m)$ is applied to the probe qubit. The amount of decoherence observed in the output of the probe qubit is a directly observable quantitative measure of the error in the estimate $A(m)$ of $\hat{A}$.
}
\end{figure}

The scenario to be considered is shown in Fig. \ref{fig1}. A probe qubit is prepared in an eigenstate of $\hat{X}$, which means that it is now maximally sensitive to phase shifts generated by the operator $\hat{Z}$. The probe qubit then interacts weakly with the target observable $\hat{A}$ of the quantum system under investigation. The interaction is described by the unitary operator
\begin{equation}
\hat{U}_{SP} = \exp\left(-i \frac{\sigma}{\hbar} \hat{A} \otimes \hat{Z} \right).
\end{equation} 
The strength of the interaction is given by the control parameter $\sigma$. In general, the interaction will be realized by applying external controls that define a time-dependent Hamiltonian. The control parameter $\sigma$ is defined by the integral of this time-dependent evolution of the interaction and its value can be controlled by the same means that are used to control the precise time dependence of the interaction. It is therefore possible to vary the strength of the interaction to ensure that only the lowest order terms in $\sigma$ contribute. 

The interaction with the system depends on the value of the property $\hat{A}$ in the input state $\hat{\rho}_S$ of the system. In general, the statistics of $\hat{A}$ can be represented by the probabilities of the eigenstates of $\hat{A}$ in the state $\hat{\rho}_S$. The uncertainty associated with this probability distribution results in decoherence, since the phase changes in the probe qubit superposition of $Z=-1$ and $Z=+1$ depend on the random eigenvalue $A_a$ of $\hat{A}$. Without any feedback, the uncompensated decoherence can be expressed by
\begin{equation}
\label{eq:random}
\langle \hat{X} \rangle(\mbox{out}) = \sum \langle a \mid \hat{\rho}_S \mid a \rangle \cos\left(2 \frac{\sigma}{\hbar} A_a \right).
\end{equation} 
For sufficiently small values of $\sigma$, the decoherence is approximately determined by the uncertainty $\Delta A^2$ of the property $\hat{A}$ in the input state $\hat{\rho}_S$. If the expectation value of $\hat{A}$ is zero,
\begin{equation}
\label{eq:uncertainty}
1-\langle \hat{X} \rangle(\mbox{out}) \approx \frac{2 \sigma^2}{\hbar^2} \Delta A^2. 
\end{equation}
The decoherence in the probe qubit thus provides us with direct evidence of the statistical fluctuations of the property $\hat{A}$ in the system state $\hat{\rho}_S$.

We can now investigate whether the outcome $m$ of a quantum measurement performed on the system after its interaction with the probe qubit contains any information about the quantity $\hat{A}$ before the measurement was performed. Since we are only interested in the relation between the measurement outcome $m$ and the quantum statistics {\it before} the measurement, it is sufficient to describe the measurement by its positive operator valued measure $\{\hat{E}(m)\}$ which summarizes the measurement interaction and the read-out of the meter in a single operator acting on the Hilbert space of the system. It is important to remember that the original system state is not available after this measurement because of the disturbance caused by the interaction. If there had not been a weak interaction between the probe qubit and the system, the measurement outcome $m$ would be the only experimental evidence of the value of $\hat{A}$ in the system and there would be no possibility of knowing how reliable an estimate $A(m)$ of $\hat{A}$ based on $m$ was. However, even a weak interaction between the system and the probe qubit creates a correlation between the phase rotation of the qubit and the value of $\hat{A}$ in the system. It is therefore possible to evaluate the precision of an estimate $A(m)$ by using the value to compensate the effects of the interaction on the probe qubit. In particular, the effects of the interaction can only be undone completely if the estimate $A(m)$ provided the correct value of $\hat{A}$. In this case, the application of a corresponding unitary operation on the probe qubit will restore maximal phase coherence to the probe. Any errors in the estimate $A(m)$ can be observed as phase fluctuations that remain after the decoherence of the qubit has been compensated by a unitary operation based on that estimate.

The experimental test of the uncertainty of the estimate of $\hat{A}$ is shown in Fig. \ref{fig1}. It consists of a feedback signal that compensates the effects of the estimated value $A(m)$ on the probe qubit by implementing the inverse of the unitary associated with this value,
\begin{equation}
\hat{U}_{Z}(m) = \exp\left(-i \frac{\sigma}{\hbar} (-A(m)) \hat{Z} \right).
\end{equation} 
We can combine this operation together with the original interaction to arrive at a more direct expression of the feedback compensation,
\begin{equation}
\hat{U}_{Z}(m) \hat{U}_{SP} = \exp\left(-i \frac{\sigma}{\hbar} (\hat{A} - A(m))\otimes \hat{Z} \right).
\end{equation} 
Since the unitary operation now depends on the measurement outcome, it is necessary to sum over all the possible outcomes $m$ to find out the net effect of the different feedback operations associated with the estimates $A(m)$. The modified expectation value of the probe qubit output is then given by
\begin{equation}
\langle \hat{X} \rangle(\mbox{out}) = \sum_m \mbox{Tr}\left((\hat{E}(m) \otimes \hat{X}) 
\hat{U}_{Z}(m) \hat{U}_{SP} (\hat{\rho}_S \otimes \hat{\rho}_{X=+1}) \hat{U}^\dagger_{SP} \hat{U}^\dagger_{Z}(m) \right).
\end{equation}
This equation is greatly simplified by the fact that the unitary operators commute with the $\hat{Z}$ operator of the probe qubit. Since the operator $\hat{X}$ exchanges the eigenstates of $\hat{Z}$, the equation is a sum of two complex conjugate terms in which opposite eigenvalues of $\hat{Z}$ appear in the unitary operations between which the input state is sandwiched. The result is an expression that refers only to the Hilbert space of the system,
\begin{equation}
\label{eq:full}
\langle \hat{X} \rangle (\mbox{out}) = \mbox{Re}\left(\sum_m \mbox{Tr}\left(\hat{E}_S(m) \exp(i \frac{\sigma}{\hbar} (\hat{A}-A(m))) \hat{\rho}_S \exp(i \frac{\sigma}{\hbar} (\hat{A}-A(m)) \right) \right).
\end{equation}
Note the similarity of this equation with Eq.(\ref{eq:random}), where the sum ran over eigenstates of $\hat{A}$. In the feedback compensated result, the sum must run over the actual measurement outcomes, and the value of $\hat{A}$ is still represented by an operator. Eq.(\ref{eq:full}) evaluates the fluctuations of $\hat{A}$ without assigning error free values of $\hat{A}$ to each outcome $m$. For sufficiently weak interactions, the decoherence that remains after feedback compensation is given by
\begin{equation}
1-\langle \hat{X} \rangle (\mbox{out}) \approx \frac{2 \sigma^2}{\hbar^2} 
\sum_m \mbox{Tr}\left(\hat{E}_S(m)(\hat{A}-A(m)) \hat{\rho}_S(\hat{A}-A(m))\right).
\end{equation}
The uncompensated decoherence represents the amount by which the estimates $A(m)$ differ from the actual phase shifts induced by the operator $\hat{A}$ in the interaction with the probe qubit. The amount of uncompensated decoherence can therefore be used to evaluate the error of the estimates $A(m)$. Comparison with Eq.(\ref{eq:uncertainty}) shows that the quantitative error corresponds to a quantum uncertainty $\epsilon_A$ of the measurement outcomes $A(m)$ given by
\begin{equation}
\label{eq:Oz}
\epsilon_A^2 = \sum_m \mbox{Tr}\left(\hat{E}_S(m)(\hat{A}-A(m)) \hat{\rho}_S(\hat{A}-A(m))\right).  
\end{equation}
Remarkably, this formula is identical to the general definition of measurement errors given by Ozawa in \cite{Oza03} even though it describes the directly observable amount of decoherence in a feedback compensation scenario. Specifically, the estimates $A(m)$ represent the measurement results associated with the measurement outcomes $m$ originally obtained by measurements of the meter in Ozawa's theory. It may be important to note that Ozawa's original formulation represents the measurement by a read-out operator acting on a meter system after a measurement interaction. Eq.(\ref{eq:Oz}) is obtained by eliminating the meter system. This is done by summing over the eigenstates $\mid m \rangle$ of the read-out in the meter system, where the operators $\hat{E}_S(m)$ represent the effects of the measurement interaction on the system conditioned by the read-out of $A(m)$ in the meter. A similar form of Ozawa uncertainties is commonly used in discussions of joint measurements and the relation between Ozawa uncertainties and weak values \cite{Lun10,Iin16,Bra13}. 

The present feedback scenario is consistent with the explanation of Ozawa uncertainties by Hall in \cite{Hal04}, where it was shown that the measurement result of Ozawa's theory corresponds to an estimate based on the initial information represented by the input state $\hat{\rho}_S$ and the final information represented by the outcomes $m$ of the measurement $\{\hat{E}(m)\}$. However, Hall's theory did not describe any experimental consequences of the errors in the estimate. It was generally assumed that discussions of measurement uncertainties had to be based on theoretical speculations regarding the unobservable error-free values of the physical property $\hat{A}$ \cite{Roz15,Ren17}. With regard to Ozawa's proposal, all previous experimental confirmations were based on theoretical arguments regarding the measurement outcomes obtained with input states different from the one for which the uncertainty was determined \cite{Iin16,Erh12,Kan14,Sul17}. It is therefore extremely important to recognize that, contrary to previous expectations, the present result shows that Ozawa uncertainties characterize an actual physical phenomenon associated with a specific combination of input state $\hat{\rho}_S$ and measurement $\{\hat{E}(m)\}$. The Ozawa uncertainty of any given measurement is experimentally observable as uncompensated decoherence following a feedback compensation based on the measurement result $A(m)$ associated with an outcome of $m$ of the measurement performed on the system after its interaction with the probe qubit. 

There are a number of very important consequences of this result relating to the previously known properties of Ozawa uncertainties \cite{Lun10,The18,Hof11}. Most importantly, the measurement results $A(m)$ assigned to the outcomes $m$ can be optimized to reduce the decoherence to its minimal value. The result of this optimization corresponds to an assignment of weak values to the measurement outcomes,
\begin{equation}
A_{\mathrm{opt.}}(m) = \mbox{Re} \left(\frac{\mbox{Tr}(\hat{E}_S(m) \hat{A} \hat{\rho}_S)}{\mbox{Tr}(\hat{E}_S(m)\hat{\rho}_S)}\right).
\end{equation}
Weak values therefore provide the best estimate for a compensation of decoherence induced by the quantum fluctuations of $\hat{A}$. Note that this result is obtained from the analysis of a feedback compensation procedure that is independent of the original definitions of weak values. The emergence of weak values in the present context shows that weak values represent empirically valid evaluations of a physical quantity between state preparation and measurement. They can be defined operationally without any interpretational assumptions about the physics of quantum measurements, revealing a serious flaw in arguments that seek to explain weak values in terms of quantum interference effects associated with weak measurements \cite{Dre15,Sok16}. In addition, the Ozawa uncertainty $\epsilon_A$ associated with the weak values $A_{\mathrm{opt.}}(m)$ provides empirical evidence of how well these weak values describe the quantity $\hat{A}$ between state preparation and measurement. As shown in previous works, the Ozawa uncertainty drops to zero for projective measurements of pure states if all of the weak values are real,
\begin{equation}
\epsilon_A=0 \hspace{0.5cm} \mbox{for} \hspace{0.5cm} A(m)= \frac{\langle m \mid \hat{A} \mid \psi \rangle}{\langle m \mid \psi \rangle}.
\end{equation}
This condition is satisfied by a wide range of projective measurements $\{ \mid m \rangle \}$, indicating that error free measurements can be very different from the conventional projections onto eigenstates of the operator $\hat{A}$. The fluctuations of $\hat{A}$ in $\mid \psi \rangle$ are not defined by the eigenvalues of $\hat{A}$ and their probabilities, but take contextual values depending on the type of measurement made to complement the information about $\hat{A}$ already available in $\mid \psi \rangle$. This observation may be of particular importance when the decoherence effects of different non-commuting observables need to be compensated at the same time. In this case the uncertainty relations given by Ozawa in \cite{Oza03} and later improved upon by Branciard in \cite{Bra13} provide the correct limit of a joint compensation protocol. The analysis of feedback compensation thus reveals how quantum contextuality works in a wide range of practically relevant situations. 

In conclusion, the analysis presented above shows that feedback compensation of decoherence caused by sufficiently weak interactions of the system with a probe qubit can be used as a direct experimental evaluation of the measurement uncertainty for the outcomes $A(m)$ assigned to each individual result $m$. The analysis of the feedback compensation scenario shows that the theory derived by Ozawa correctly describes the amount of uncompensated decoherence. Different from previous experimental tests of Ozawa uncertainties \cite{Iin16,Erh12,Kan14,Sul17}, no tomographic reconstructions are needed and the experimental evidence permits no other interpretation except that the weak force exerted by the system on the probe qubit fluctuates with the value given by the Ozawa uncertainties. In addition, the optimal feedback is obtained when the weak values associated with each measurement result are used as outcomes of the measurement, confirming weak values as the optimal estimates of physical properties between preparation and measurement. In the pure state limit where the Ozawa uncertainties drop to zero, weak values provide an accurate description of the contextual quantum fluctuations of an observable $\hat{A}$ \cite{Lun10,The18,Hof11}. Feedback compensation of decoherence thus confirms the empirical validity of both Ozawa's generalization of uncertainty and of weak values without requiring any untestable assumptions. 


\end{document}